\documentclass[conference]{IEEEtran}

\usepackage{amsmath,amssymb,amsfonts}
\usepackage{algorithmic}
\usepackage{graphicx}
\usepackage{textcomp}
\usepackage{xcolor}
\usepackage[numbers]{natbib}
\usepackage{{booktabs}}
\usepackage{comment}
\usepackage{colortbl}
\usepackage{amsfonts}
\usepackage[hyphens]{url}
\usepackage{hyperref}
\hypersetup{hidelinks}
\usepackage{longtable}
\usepackage[yyyymmdd]{datetime}
\usepackage{verbatim}  
\usepackage{array}     
\usepackage{mathtools} 
\usepackage{amsmath}
\usepackage{makecell}
\usepackage{balance}
\usepackage[capitalise]{cleveref}
\usepackage[textwidth=2.7cm,textsize=scriptsize]{todonotes}
\usepackage{soul}

\def\BibTeX{{\rm B\kern-.05em{\sc i\kern-.025em b}\kern-.08em
    T\kern-.1667em\lower.7ex\hbox{E}\kern-.125emX}}

\begin{document}

\title{The Role of Legacy Mobile Networks in Infrastructure Resilience: Evidence from the Southern Brazil Flood}

\author{\IEEEauthorblockN{\textsuperscript{} Daniel Meyer, Lisandro Z Granville, Leandro M. Bertholdo}
\IEEEauthorblockA{\textit{Institute of Informatics - Federal University of Rio Grande do Sul}} 
\textit{Av. Bento Gonçalves, 9500 – Porto Alegre, Brazil}\\

{(daniel.meyer, leandro.bertholdo)}@ufrgs.br, granville@inf.ufrgs.br}

\maketitle
\renewcommand{\thefootnote}{} 
\footnotetext{\textbf{Preprint: This is the version submitted to IEEE GLOBECOM 2025 (before peer review). The Final Accepted Version will be published in IEEE Xplore.}}
\renewcommand{\thefootnote}{\arabic{footnote}}

\begin{abstract}
This paper investigates the resilience of mobile communication networks during the extreme flooding that affected Rio Grande do Sul, Brazil, in May 2024. Based on regulatory data and technical insights from operators, the study identifies the main causes of network disruptions, primarily related to flooding and prolonged power outages. The results reveal the significant vulnerability of modern networks (4G/5G) during the event and the essential role played by legacy technologies (2G/3G) in sustaining basic connectivity under adverse conditions. The findings highlight the need for disaster-aware infrastructure planning, considering the continued importance of legacy systems, diversified power supply strategies, and resilient network designs to improve service continuity during future crises.
\end{abstract}


\section{Introduction}

In May 2024, Brazil’s southernmost state, Rio Grande do Sul, experienced an unprecedented climatic disaster that disrupted critical infrastructure and displaced approximately 2.3 million people across more than 400 municipalities. The metropolitan region of Porto Alegre, the state capital, faced the most severe impacts, with parts of the city submerged for over 30 consecutive days. Among the most affected services were mobile communication networks, essential for emergency response and public safety. The prolonged flooding severely impacted telecommunications, leading to widespread outages of Base Transceiver Stations (BTS) and significant degradation of service continuity. In the most critical moments, even basic communication services were compromised.

To mitigate the impacts, telecom operators adopted emergency measures such as free-roaming agreements. Nevertheless, modern networks (4G and 5G) suffered substantial outages, whereas legacy technologies (2G and 3G)---although limited in capacity---remained operational and played a vital role in sustaining communication under adverse conditions. 

This event exposed the vulnerability of mobile networks to long-duration disasters and raised critical questions regarding the planned decommissioning of older technologies. Before the flood, the Brazilian National Telecommunications Agency (Anatel) had already initiated efforts to sunset 2G and 3G networks, aligned with international modernization trends~\cite{Anatel2023Transicao2G}. However, the dependence on legacy technologies during the disaster demonstrated the potential risks associated with this transition.

In this context, this paper investigates two research questions: (i) What were the main factors leading to mobile service disruptions during the 2024 flood? (ii) How would mobile coverage have been affected if 2G and 3G had already been phased out?

To address these questions, we combine regulatory datasets, operator network snapshots, and interviews with telecommunications professionals. The results provide insights into the root causes of network failures and demonstrate the essential role of legacy technologies in supporting mobile network resilience during climate-induced disasters.

\section{Related Work}
\label{related}

The resilience of telecommunications infrastructure has been extensively studied, focusing on economic impacts, network performance, and recovery strategies~\cite{JAHN201529}. However, most studies address short-duration climatic events, while the 2024 flooding in Brazil stands out as a long-lasting disaster, severely affecting Porto Alegre and surrounding municipalities~\cite{Temporaisrscronologia}.

Several case studies highlight the vulnerability of telecommunication systems to extreme weather events. Hurricane Katrina (2005) caused extensive damage along the Gulf Coast of Mexico, leaving 70 percent of cellular base stations inoperative due to strong winds and flooding, delaying full service restoration by up to 40 days~\cite{kwasinski2009telecommunications}. Although Porto Alegre did not experience destructive winds, the slow-rising flood resulted in similarly prolonged service disruptions.

Hurricanes Irma and Maria (2017) devastated Puerto Rico, destroying 91 percent of its telecommunications infrastructure and exposing the fragility of aerial networks~\cite{CORDOVA2021102106}. This vulnerability is especially relevant to Brazil, where approximately 99 percent of the telecommunications infrastructure is deployed above ground, contrasting with countries such as Spain (25 percent) and the Netherlands, where networks are fully underground~\cite{Martins2022lise}.

In October 2024, Hurricane Milton disrupted communication services in Florida, impacting emergency response operations~\cite{EUAFuracao}. Similarly, floods in Valencia, Spain, affected more than 200,000 users, with 68 percent of mobile lines restored within one week and nearly full recovery achieved in two weeks~\cite{Laslluvias}. In Rio Grande do Sul, recovery was significantly slower, taking over 120 days to restore most services.

In previous work~\cite{bertholdo2025analyzing}, we conducted a broader assessment of this disaster's impact on the state's information and communication infrastructure, focusing on optical networks, data centers, and Internet exchanges. However, the specific resilience and performance of mobile networks, which play a critical role during emergencies, have not been thoroughly examined. This paper addresses this gap by investigating the behavior of mobile infrastructure during the 2024 flood, identifying failure causes, and analyzing the role of legacy technologies in maintaining connectivity.

\section{Datasets}
\label{datasets}

This study employs datasets from Anatel, including information on Personal Mobile Service (SMP) stations, such as location, supported technologies, frequency allocations, and network impact records~\cite{Redes_Anatel}. Inconsistencies were identified, notably in the coordinates of several Base Transceiver Stations (BTS), which were corrected using supplementary data provided by a mobile operator in the region.

Additional insights were obtained through consultations with professionals from two mobile operators, including technicians, analysts, supervisors, engineers, and specialists. These interactions provided information on operational conditions, maintenance constraints, and field measurements collected by technician teams. The collected measurements contributed to understanding mobile network performance under adverse conditions in dense urban areas. The flood extent was determined using georeferenced maps from the Hydraulic Research Institute (IPH/UFRGS), corresponding to May 6, 2024, when Lake Guaíba reached its peak level of 5.35 meters~\cite{Temporaisrscronologia}.

\section{Methodology}
\label{methodology}

This section describes the methods employed to address the two research questions outlined in this work.

\subsection{Analysis of Factors Contributing to Service Disruption}

Natural disasters, such as storms and floods, commonly affect telecommunications systems through fiber cuts, antenna failures, equipment malfunctions, vandalism, and power outages. 

To investigate the causes of mobile service disruption during the 2024 flood, we combined Anatel’s data, operator network records, and the field measurements and observations collected during the emergency. These datasets enabled the identification of affected BTS sites, classification of outage causes, and correlation of failures with infrastructure vulnerability and flood extent.

Mobile sites were classified as affected if their locations intersected the flood polygons provided by IPH/UFRGS and were concurrently reported as offline in the operator's management system. Accessibility constraints were identified based on field reports indicating areas where maintenance activities could not be performed during the flood.

\subsection{Analysis of the Impact of 2G and 3G Deactivation}

In the affected regions, particularly the Porto Alegre Metropolitan area, 2G and 3G networks contributed to maintaining minimum service continuity during the flood. Despite their relevance under such conditions, Anatel plans to phase out these legacy technologies in favor of 4G and 5G~\cite{Anatel2023Transicao2G}, although they still account for over 20\% of mobile connections in Brazil. While modernization improves network performance, it may reduce resilience under adverse conditions~\cite{GSMA-emergency}.

To evaluate the potential impact of legacy network decommissioning, a comparative coverage analysis for 2G, 3G, 4G, and 5G was conducted within the flooded regions of Porto Alegre. Each BTS was associated with its nominal coverage radius, defined by its supported technologies, assuming ideal propagation conditions.

\section{Limitations}
\label{limitations}

This study faced two main limitations. First, access to detailed operational data was constrained by compliance and confidentiality restrictions, limiting the granularity of the analysis. Second, spatial analyses were performed using the WGS84 coordinate system (EPSG:4326) without projection to a metric system or modeling terrain and building-related signal losses. These simplifications may have introduced minor distortions in coverage and distance estimations. Future work will address these limitations by adopting projected coordinate systems and fostering partnerships to improve data availability.

\section{Results}
\label{results}

This section presents the results obtained from the analysis of: 
(i) the factors contributing to mobile network disruptions during the 2024 flood, and 
(ii) the role of legacy technologies in sustaining coverage under adverse conditions.

\subsection{Causes of Mobile Network Disruption}

\begin{figure}[ht]
\centering
\includegraphics[width=1\linewidth]{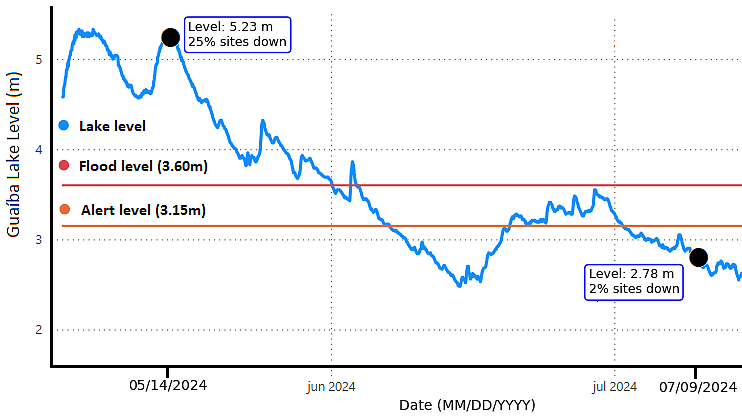}
\caption{Evolution of the Lake Guaíba water level in Porto Alegre during the 2024 flood. Snapshots corresponding to the critical (May 14) and recovery (July 9) phases are highlighted. A more complete timeline of events is available in the datasets presented in~\cite{bertholdo2025analyzing}.}
\label{fig:Nivel_gub}
\end{figure}

The evolution of Lake Guaíba's water level is shown in \autoref{fig:Nivel_gub}, providing a temporal reference for the two key snapshots analyzed in this study: the peak disruption phase (May 14, 2024) and the recovery phase (July 9, 2024). These moments were selected as representative of the most critical and recovery periods, respectively.

The flood had a direct and prolonged impact on telecommunications infrastructure, primarily due to water-induced equipment damage and power outages lasting nearly 30 days~\cite{Chuvas_RS_internet}. Despite multiple requests, the power utility did not provide detailed information on the geographic extent or duration of the outages.

To quantify the impact, two datasets collected during the event's peak (May 5, 2024) were analyzed. The first, provided by Anatel, aggregated information from all mobile operators in Rio Grande do Sul and indicated that 263 out of the 497 municipalities (53\%) experienced disruptions~\cite{Redes_Anatel}. The second dataset, obtained from a mobile operator, offered a detailed view of site-level disruptions affecting its infrastructure.

The operator snapshot from May 14, 2024, corresponding to the flood’s peak water level, revealed severe degradation, with 76 out of 305 sites (approximately 25\%) out of service in the capital Porto Alegre, affecting 1,356 BTSs. Outage causes were classified as shown in \autoref{tab:Sites-fora}: water damage (31.6\%), power outages (27.6\%), and site inaccessibility (25\%), the latter referring to locations where technicians could not perform inspections.

\begin{table}[ht]
\centering
\footnotesize	
\caption{Classification of out-of-service sites in Porto Alegre during the critical phase (May 14, 2024).}
\label{tab:Sites-fora}
\begin{tabular}{|p{3.4cm}|c|c|}
\hline
\textbf{Cause} & \textbf{Quantity} & \textbf{Percentage (\%)} \\ \hline
Flooded Site               & 24              & 31.6        \\ \hline
No Power Supply            & 21              & 27.6        \\ \hline
No Site Access             & 19              & 25.0        \\ \hline
No Information Available     & 9               & 11.8        \\ \hline
Fiber Ring Break           & 2               & 2.6         \\ \hline
Vandalism                  & 1               & 1.3         \\ \hline
\textbf{Total}             & \textbf{76}     & \textbf{100} \\ \hline
\end{tabular}
\end{table}

The mobile operator's network comprised (5.16\%) of BTSs operating on 2G, (13.97\%) on 3G, (72.76\%) on 4G, and only (8.11\%) on 5G.

The 2G infrastructure assessment revealed that 60 BTSs (22.06\%) were out of service at the peak of the flooding, while (77.94\%) remained operational.

For the 5G infrastructure, the analysis showed that 109 BTSs (25.47\%) went offline, while 315 BTSs (73.60\%) continued operating and 4 BTSs . A small fraction (0.93\%) had an undefined status due to incomplete management data.

As shown in \autoref{tab:failure_rate}, the failure rates varied across technologies. Notably, 2G BTSs exhibited the lowest failure rate (22.0\%) compared to newer technologies, despite their lower network share. This finding suggests that legacy technologies contributed significantly to network resilience during the event.

\begin{table}[ht!]
\centering
\footnotesize
\caption{Failure Rate by Mobile Technology during the Flood.}
\label{tab:failure_rate}
\begin{tabular}{|>{\centering\arraybackslash}p{1.3cm}|>{\centering\arraybackslash}p{1.8cm}|>{\centering\arraybackslash}p{1.8cm}|>{\centering\arraybackslash}p{2.2cm}|>{\centering\arraybackslash}p{2cm}|}
\hline
\textbf{Technology} & \textbf{Total BTSs} & \textbf{Affected BTSs} & \textbf{Failure Rate (\%)}\\ \hline
2G & 272            & 60      & 22.0                 \\ \hline
3G & 737          & 200      & 27.1                   \\ \hline
4G & 3838            & 987       & 25.7                 \\ \hline
5G & 428             & 109     & 25.4               \\ \hline
\textbf{Total} & \textbf{5,275} & \textbf{1,356}  & \textbf{25.7} \\ \hline

\end{tabular}
\end{table}

By July 9, 2024, after the water level had dropped below the alert threshold of 3.15 meters, only seven sites (2\%) remained inoperative, indicating that the network had almost fully recovered. However, the operator did not clarify whether these sites were pending restoration or had been permanently decommissioned, citing confidentiality restrictions.

\subsection{Service Layer Impact: Internet Access Disruption}

\begin{figure}[htbp]
\centerline{\includegraphics[width=1\linewidth]{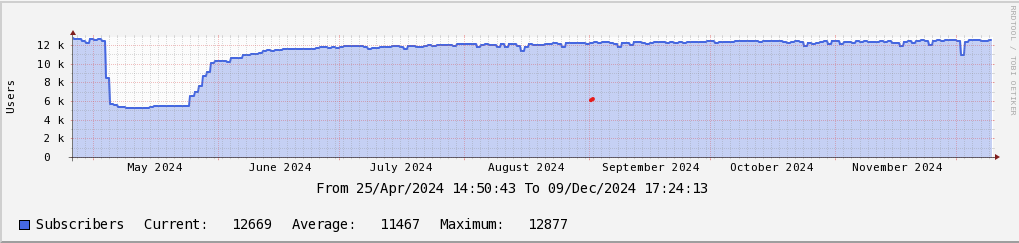}}
\caption{Internet user count connected to a traffic concentrator in Porto Alegre (May--November 2024). A sudden reduction of nearly 50\% occurred during the May flood, followed by a gradual recovery.}
\label{fig:Concentrador}
\end{figure}

Beyond infrastructure-level outages, the flood caused substantial degradation at the service layer, directly impacting user connectivity. \autoref{fig:Concentrador} presents the evolution of the number of Internet users connected to one of the main traffic concentrators in Porto Alegre between May and November 2024. During the peak of the flood, a sharp drop of approximately 50\% in active users was observed, reflecting the combined effects of site outages, prolonged power failures, and the collapse of local coverage.

Complementary evidence was obtained from end-user measurements collected by a nation-wide network quality monitoring system~\cite{bertholdo2025analyzing}. 
\autoref{tab:simet_mobile} summarizes the evolution of median Bandwidth (BW) and Round Trip Time (RTT) for mobile network users in the most affected cities. The data confirm a significant degradation in perceived quality: the median mobile bandwidth dropped from 119.5 Mbps before the event to 101.2 Mbps during the critical phase, and further to 69.5 Mbps during the recovery period. Meanwhile, median RTT temporarily decreased during the flood, likely due to reduced user demand, but worsened in the aftermath, suggesting that congestion and capacity limitations persisted even after partial network restoration.

\begin{table}[ht]
    \centering
    \caption{Mobile network performance in the most affected cities.}
    \label{tab:simet_mobile}
    \footnotesize
       \begin{tabular}{|>{\centering\arraybackslash}p{1.6cm}|>{\centering\arraybackslash}p{2.5cm}|>{\centering\arraybackslash}p{1.6cm}|>{\centering\arraybackslash}p{1.2cm}|}
        \hline
        \textbf{Period} &\textbf{Measurements from the total (\%)} &\textbf{Median BW (Mbps)} & \textbf{Median RTT (ms)} \\ \hline
        Pre-event &33.5 &119.5 & 28.6 \\ \hline
        Critical days & 22.0&101.2 & 23.7 \\ \hline
        Post-event & 44.5&69.5 & 36.7 \\ \hline
    \end{tabular}
\end{table}

Taken together, the operator-side data (concentrator user counts) and end-user measurements bandwidth and RTT consistently show that the 2024 flood caused not only infrastructure failures but also a severe deterioration of service quality, even in areas where connectivity remained technically available.

These findings illustrate the progressive degradation and partial recovery of mobile network infrastructure in Porto Alegre, affecting both infrastructure availability and service quality. The combined evidence from site outages and end-user measurements highlights the limitations of the current mobile network under extreme events.

\subsection{Impact of Legacy Networks on Coverage}

To assess the role of legacy networks during the flood, a comparative analysis was conducted between 2G and 5G coverage, focusing on their ability to maintain communication in the affected areas. Some of their technical characteristics are presented in\autoref{tab:tecnologias-redes}. Individually, these values are influenced by factors such as obstacles, terrain morphology, antenna height, and the maximum transmission power permitted by regulatory agencies. 

\begin{table}[ht!]
\centering
\footnotesize
\caption{Coverage characteristics of mobile technologies used in the analysis.}
\label{tab:tecnologias-redes}
\begin{tabular}{|>{\centering\arraybackslash}p{1.3cm}|>{\centering\arraybackslash}p{1.13cm}|>{\centering\arraybackslash}p{1.5cm}|>{\centering\arraybackslash}p{1.5cm}|>{\centering\arraybackslash}p{1.13cm}|}


\hline
\textbf{Technology} & \textbf{Coding} & \textbf{Frequency (MHz)} & \textbf{Bandwidth (MHz)} & \textbf{Range (km)} \\ \hline
2G & GSM            & 850 / 1800      & 2.5 / 5         & 2 -- 10         \\ \hline
3G & WCDMA          & 850 / 2100      & 10 / 20         & 1 -- 5          \\ \hline
4G & LTE            & 700--2600       & 10 -- 20        & 1 -- 4          \\ \hline
5G & NR             & 2300 / 3500     & 50 / 100        & 0.5 -- 1        \\ \hline
\end{tabular}
\end{table}

Although channel bandwidth does not directly determine the signal propagation distance, it significantly affects link capacity. For a given modulation scheme and equivalent radio frequency (RF) conditions, wider bandwidth channels enable higher connection speeds and support a greater number of simultaneous users.

 Modulation schemes dynamically adapt according to propagation conditions: near the site and under line-of-sight conditions, higher-order modulations are typically used, while in distant areas with more attenuated RF signals, lower-order modulations are employed~\cite{9486300}.

The coverage ranges adopted in this study were based on practical values calibrated through field measurements obtained via WalkTest and DriveTest procedures~\cite{6211483} . While WalkTest is conducted on foot—typically in indoor environments or densely urbanized areas—DriveTest uses equipment installed in vehicles to cover larger distances. Both methods provide essential data for network performance assessment and fault identification.

\begin{figure}[ht!]
\centering
\includegraphics[width=1\linewidth]{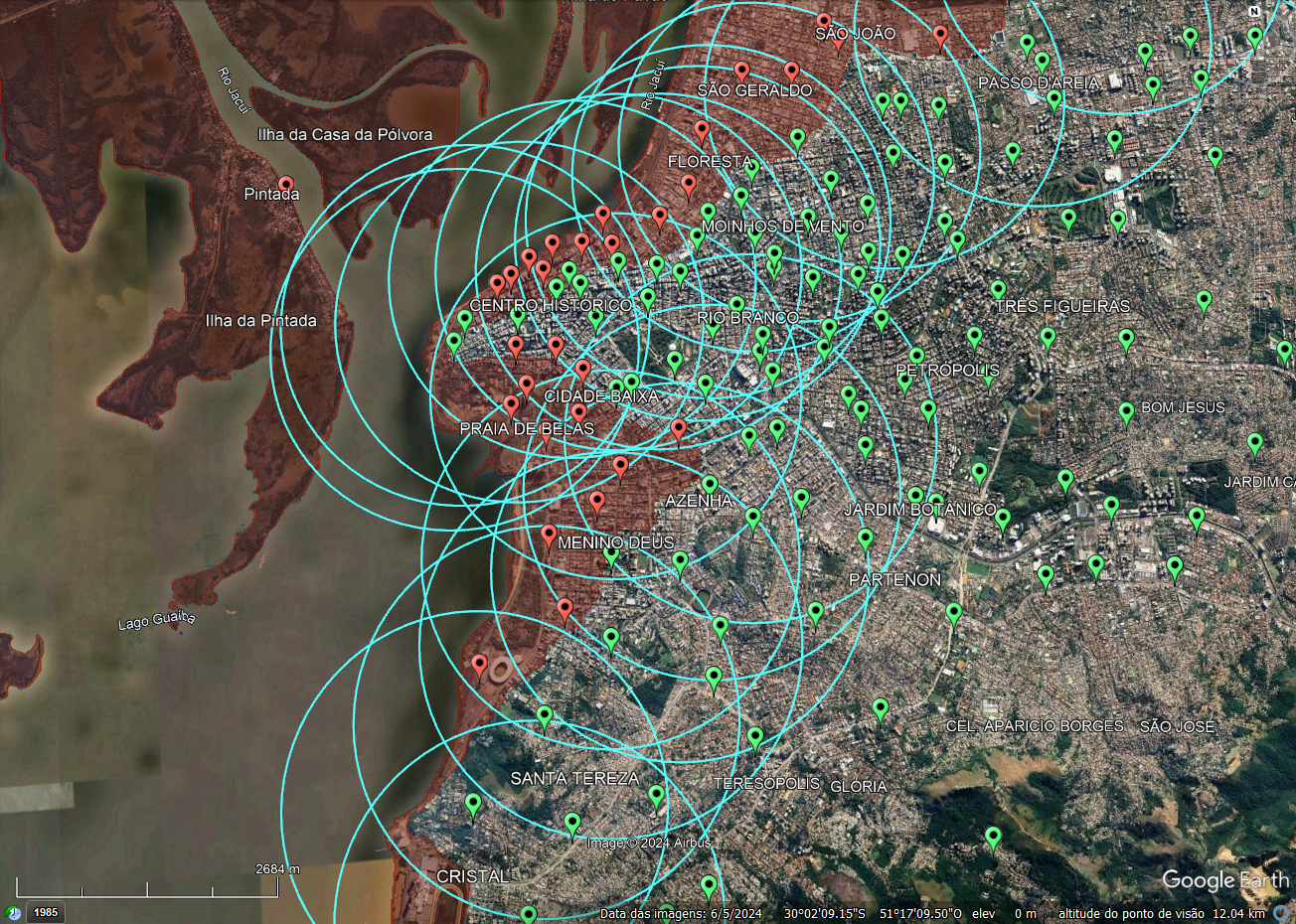}
\caption{Estimated 2G coverage maps of a mobile operator's stations in the city of Porto Alegre - (2km radius).}

\label{fig:Mapa2g}
\end{figure}

\begin{figure}[ht!]
\centering
\includegraphics[width=1\linewidth]{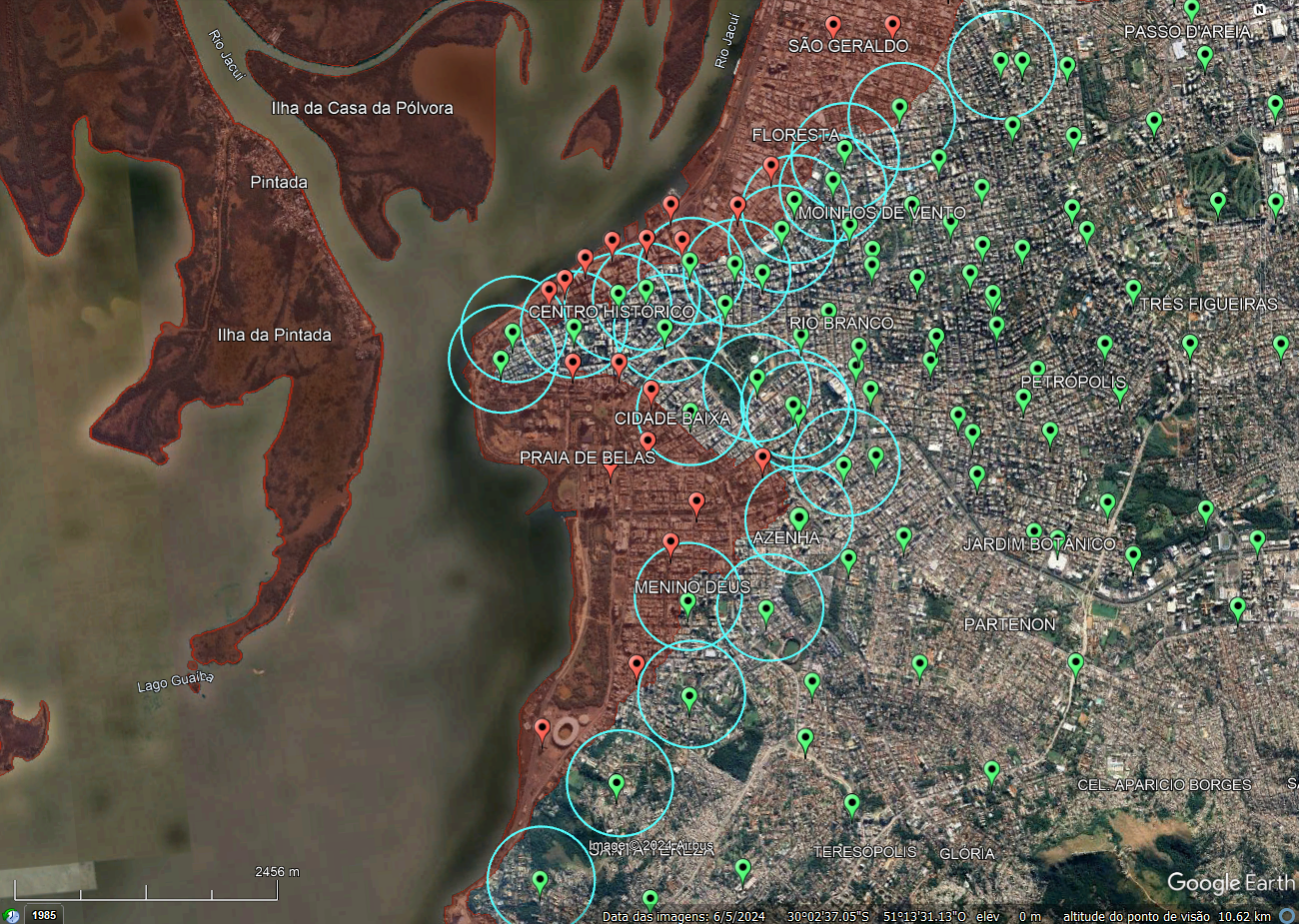}
\caption{Estimated 5G coverage maps of a mobile operator's stations in the city of Porto Alegre - (0.5km radius).}

\label{fig:Mapa5g}
\end{figure}

For the map generation shown in Figures~\ref{fig:Mapa2g} and ~\ref{fig:Mapa5g}, we simulated a worst-case signal coverage scenario, adopting the minimum coverage radius for each technology: 2 km for 2G and 0.5 km for 5G, as specified in\autoref{tab:tecnologias-redes}. These maps represent the estimated mobile coverage in Porto Alegre on May 6, 2024, combining flood extent data—corresponding to the peak water level of 5.35 meters~\cite{Mapa_interativo_RioGrande}—with the technological composition and spatial distribution of each Base Transceiver Station. The data on location, frequency, and supported technology for each BTS was provided by the mobile operator.

The red-shaded polygon delineates the flooded area, while the blue circles represent the estimated coverage zones of each active Base Transceiver Station located at the flood boundary, according to the technology in use (2G and 5G). Red markers identify the sites that went out of operation during the event, whereas green markers indicate the sites that remained operational.

This comparison highlights the crucial role of legacy technologies in extending coverage during emergencies. While modern networks such as 4G and 5G prioritize data capacity and spectral efficiency, their reduced signal range, especially at higher frequencies, limits their resilience under adverse conditions. In contrast, 2G and 3G networks, originally designed for voice and basic data services, offer wider coverage and superior signal penetration, which proved essential during the flood~\cite{9512046}.

The analysis reveals a significant disparity in coverage between 2G and 5G, with 2G demonstrating broader signal reach in the flood-affected areas due to its lower operational frequencies. Notably, even in the most severely affected zones, 2G cells remained operational in higher proportion compared to 5G, ensuring basic voice and SMS connectivity in the absence of modern data services.

This comparison highlights the crucial role of legacy technologies in extending coverage during emergency situations. While modern networks such as 4G and 5G prioritize data capacity and spectral efficiency, their reduced coverage—especially at higher frequencies—limits their resilience under adverse conditions. In contrast, 2G and 3G networks, originally designed for voice and basic data services, offer broader coverage and superior signal penetration, which proved essential during the flood~\cite{9512046}.

These findings reinforce the importance of legacy technologies during disasters and point to the need for implementing 5G and future technologies in lower frequency bands, such as 700 MHz, to enable greater signal reach and network availability—particularly in areas with lower infrastructure density. Such measures would enhance overall network resilience, especially in regions prone to extreme weather events.

\section{Discussion}

The analysis revealed that approximately one-quarter of the mobile sites in Porto Alegre became inoperative during the flood's critical phase, primarily due to water-induced equipment failures and power outages. Additionally, around 25\% of the affected sites could not be assessed due to accessibility constraints, further limiting operators' ability to deploy mitigation measures.

This disruption occurred under an unprecedented hydrological scenario. Historically, between 1899 and 2023, Lake Guaíba surpassed the 3-meter level only four times. However, between September 2023 and May 2024, it exceeded this threshold on three occasions, culminating in the record-breaking flood of 2024~\cite{moraes2024}. This trend underscores the growing frequency and intensity of extreme weather events, aligning with global climate change projections.

Although free roaming agreements between carriers have been activated to mitigate the degradation of service, connectivity continuity depended largely on legacy technologies, particularly 2G. Modern networks such as 5G were most affected due to the smaller coverage range. Coverage analysis showed that if only modern technologies had been available, the impact on users would have been significantly worse.

Beyond technical aspects, the event exposed critical limitations in both energy backup strategies and infrastructure placement. Operators implemented lithium batteries and portable generators to sustain essential sites, but these solutions proved insufficient to endure a power outage lasting nearly 30 days. This highlights the need to reassess backup strategies and to consider deploying alternative energy systems that combine multiple sources, such as solar, wind, and fuel-based generators, to ensure service continuity during prolonged crises. Likewise, the dependence on aerial cabling left critical infrastructure vulnerable to physical damage and hindered field teams' access for repairs during the flood. 

Recent regulatory efforts, such as Municipal Law No. 13.402~\cite{LEI_13402}, mandating the underground installation of new utility networks in Porto Alegre, represent an important step toward improving infrastructure resilience. However, such transitions require long-term investments and coordinated efforts between public authorities and private operators.

Finally, the adoption of network softwarization and virtualization approaches, such as OpenRAN~\cite{OpenRan}, could further strengthen resilience by enabling more flexible orchestration, facilitating the rapid deployment of temporary infrastructure, and reducing dependency on single points of failure. Combined with improved documentation practices and better data-sharing frameworks, these strategies could help prepare communication networks for the growing threat of long-duration disasters.

\section{Conclusion}

This study investigated the resilience of mobile communication networks during the 2024 flood in Porto Alegre, revealing critical vulnerabilities in both infrastructure and service layers. Legacy technologies (2G/3G) proved essential for maintaining minimal connectivity when modern networks (4G/5G) were severely impacted by equipment damage and long-lasting power outages. However, relying on outdated technologies is no longer viable as they are being progressively decommissioned worldwide.

The results emphasize the urgency of adopting disaster-resilient network architectures, combining robust power backup systems, diversified transmission media, and the flexible deployment of modern technologies. In particular, expanding 5G in lower-frequency bands and improving network densification will be crucial to sustaining communication during future extreme events.

Equally important is the need to establish collaborative frameworks that allow for secure and ethical sharing of operational data between operators, researchers, and public authorities. Such cooperation is fundamental to designing evidence-based strategies that enhance the resilience of critical communication services.

Strengthening mobile infrastructure against climate-induced disruptions is not optional but necessary. The increasing frequency and severity of such events call for immediate action to ensure that essential services remain operational when most needed.

\balance

\bibliographystyle{IEEEtranN}
\bibliography{refs}

\begin{thebibliography}{19}
\providecommand{\natexlab}[1]{#1}
\providecommand{\url}[1]{#1}
\csname url@samestyle\endcsname
\providecommand{\newblock}{\relax}
\providecommand{\bibinfo}[2]{#2}
\providecommand{\BIBentrySTDinterwordspacing}{\spaceskip=0pt\relax}
\providecommand{\BIBentryALTinterwordstretchfactor}{4}
\providecommand{\BIBentryALTinterwordspacing}{\spaceskip=\fontdimen2\font plus
\BIBentryALTinterwordstretchfactor\fontdimen3\font minus
  \fontdimen4\font\relax}
\providecommand{\BIBforeignlanguage}[2]{{%
\expandafter\ifx\csname l@#1\endcsname\relax
\typeout{** WARNING: IEEEtranN.bst: No hyphenation pattern has been}%
\typeout{** loaded for the language `#1'. Using the pattern for}%
\typeout{** the default language instead.}%
\else
\language=\csname l@#1\endcsname
\fi
#2}}
\providecommand{\BIBdecl}{\relax}
\BIBdecl

\bibitem[Anatel(2023)]{Anatel2023Transicao2G}
Anatel, ``Anatel promove tomada de subsídios para transição dos padrões
  2{G} e 3{G} para 4{G} e 5{G},''
  \url{https://www.gov.br/anatel/pt-br/assuntos/noticias/anatel-promove-tomada-de-subsidios-para-transicao-dos-padroes-2g-e-3g-para-4g-e-5g},
  Oct 2023.

\bibitem[Jahn(2015)]{JAHN201529}
M.~Jahn, ``Economics of extreme weather events: Terminology and regional impact
  models,'' \emph{Weather and Climate Extremes}, vol.~10, pp. 29--39, 2015.

\bibitem[G1(2024)]{Temporaisrscronologia}
G1, ``Temporais no {RS}: veja cronologia de desastre que matou 75 pessoas,''
  \url{https://g1.globo.com/rs/rio-grande-do-sul/noticia/2024/05/05/temporais-no-rs-veja-cronologia-de-desastre.ghtml},
  May 2024.

\bibitem[Kwasinski et~al.(2009)Kwasinski, Weaver, Chapman, and
  Krein]{kwasinski2009telecommunications}
A.~Kwasinski, W.~W. Weaver, P.~L. Chapman, and P.~T. Krein,
  ``Telecommunications power plant damage assessment for hurricane
  katrina--site survey and follow-up results,'' \emph{IEEE Systems Journal},
  vol.~3, no.~3, pp. 277--287, 2009.

\bibitem[Cordova and Stanley(2021)]{CORDOVA2021102106}
A.~Cordova and K.~D. Stanley, ``Public-private partnership for building a
  resilient broadband infrastructure in puerto rico,'' \emph{Telecommunications
  Policy}, vol.~45, no.~4, p. 102106, 2021.

\bibitem[de~Abreu~Martins et~al.(2022)de~Abreu~Martins, Arango, and
  Kubota]{Martins2022lise}
\BIBentryALTinterwordspacing
B.~de~Abreu~Martins, L.~G. Arango, and L.~C. Kubota,
  ``\BIBforeignlanguage{por}{Analise sobre o enterramento de infraestrutura de
  redes dos setores de distribuicao de energia e telecomunicacoes},'' IPEA,
  Brasilia, Texto para Discussao 2727, 2022. [Online]. Available:
  \url{https://hdl.handle.net/10419/261042}
\BIBentrySTDinterwordspacing

\bibitem[Euronews(2024)]{EUAFuracao}
\BIBentryALTinterwordspacing
Euronews, ``{Over 2.6 million without power as hurricane Milton slams into
  Florida},'' Oct 2024. [Online]. Available:
  \url{https://www.euronews.com/2024/10/10/hurricane-milton-makes-landfall-in-florida}
\BIBentrySTDinterwordspacing

\bibitem[Muñoz(2024)]{Laslluvias}
R.~Muñoz, ``Las lluvias provocan numerosos cortes en la telefonía móvil e
  internet | economía | {EL PAÍS},''
  \url{https://elpais.com/economia/2024-10-30/las-lluvias-provocan-numerosos-cortes-en-la-telefonia-movil-e-internet.html},
  Oct 2024.

\bibitem[Bertholdo et~al.(2025)Bertholdo, Paredes, de~Lima~Marin, Loureiro,
  Kashiwakura, and de~Botelho~Marcos]{bertholdo2025analyzing}
L.~M. Bertholdo, R.~B. Paredes, G.~de~Lima~Marin, C.~A. Loureiro, M.~K.
  Kashiwakura, and P.~de~Botelho~Marcos, ``Analyzing the effect of an extreme
  weather event on telecommunications and information technology: Insights from
  30 days of flooding,'' in \emph{International Conference on Passive and
  Active Network Measurement}.\hskip 1em plus 0.5em minus 0.4em\relax Springer,
  2025, pp. 276--304.

\bibitem[{Anatel}(2024)]{Redes_Anatel}
{Anatel}, ``{Recuperação das Redes},''
  \url{https://informacoes.anatel.gov.br/paineis/utilidade-publica/recuperacao-das-redes},
  Jul 2024.

\bibitem[GSMA(2022)]{GSMA-emergency}
GSMA, ``{GSMA Official Document},''
  \url{https://www.gsma.com/newsroom/wp-content/uploads//NG.136-Emergency-Services-White-Paper-v1.0.pdf},
  Nov 2022.

\bibitem[Lima(2024)]{Chuvas_RS_internet}
C.~Lima, ``Chuvas no {RS} deixam 13\% dos clientes de fibra óptica da {V}ivo
  sem internet,''
  \url{https://www.minhaoperadora.com.br/2024/05/chuvas-no-rs-deixam-13-dos-clientes-de-fibra-optica-da-vivo-sem-internet.html},
  May 2024.

\bibitem[Sanchez(2021)]{9486300}
J.~M. Sanchez, ``Mobile revolution: From 2g to 5g,'' in \emph{2021 IEEE
  Colombian Conference on Communications and Computing (COLCOM)}, 2021, pp.
  1--6.

\bibitem[Hapsari et~al.(2012)Hapsari, Umesh, Iwamura, Tomala, Gyula, and
  Sebire]{6211483}
W.~A. Hapsari, A.~Umesh, M.~Iwamura, M.~Tomala, B.~Gyula, and B.~Sebire,
  ``{Minimization of drive tests solution in 3GPP},'' \emph{IEEE Communications
  Magazine}, vol.~50, no.~6, pp. 28--36, 2012.

\bibitem[UFRGS(2024)]{Mapa_interativo_RioGrande}
UFRGS, ``{Base de dados e informações geográficas na Região Hidrográfica
  do Lago Guaíba e na Lagoa dos Patos em 2024},''
  \url{https://storymaps.arcgis.com/stories/a81d69f4bccf42989609e3fe64d8ef48},
  May 2024.

\bibitem[Shah et~al.(2021)Shah, Qasim, Karabulut, Ilhan, and Islam]{9512046}
A.~F. M.~S. Shah, A.~N. Qasim, M.~A. Karabulut, H.~Ilhan, and M.~B. Islam,
  ``Survey and performance evaluation of multiple access schemes for
  next-generation wireless communication systems,'' \emph{IEEE Access}, vol.~9,
  pp. 113\,428--113\,442, 2021.

\bibitem[Moraes et~al.(2024)Moraes, Collischonn, Buffon, and
  Eckhardt]{moraes2024}
S.~R. Moraes, W.~Collischonn, F.~T. Buffon, and R.~R. Eckhardt, ``{Revisão e
  consolidação da série histórica dos níveis das cheias do rio Taquari em
  Lajeado de 1939 a 2023},'' UFRGS, UNIVATES, CPRM, Tech. Rep., 2024.

\bibitem[CMPA(2023)]{LEI_13402}
CMPA, ``Lei nº 13.402,''
  \url{https://www.camarapoa.rs.gov.br/draco/processos/137070/Lei_13402.pdf},
  Mar 2023.

\bibitem[Ericsson()]{OpenRan}
\BIBentryALTinterwordspacing
Ericsson, ``{Explore Open RAN: innovation and flexibility}.'' [Online].
  Available:
  \url{https://www.ericsson.com/en/openness-innovation/open-ran-explained}
\BIBentrySTDinterwordspacing

\end{thebibliography}


\begin{thebibliography}{00}
\bibitem{b1} G. Eason, B. Noble, and I. N. Sneddon, ``On certain integrals of Lipschitz-Hankel type involving products of Bessel functions,'' Phil. Trans. Roy. Soc. London, vol. A247, pp. 529--551, April 1955.
\bibitem{b2} J. Clerk Maxwell, A Treatise on Electricity and Magnetism, 3rd ed., vol. 2. Oxford: Clarendon, 1892, pp.68--73.
\bibitem{b3} I. S. Jacobs and C. P. Bean, ``Fine particles, thin films and exchange anisotropy,'' in Magnetism, vol. III, G. T. Rado and H. Suhl, Eds. New York: Academic, 1963, pp. 271--350.
\bibitem{b4} K. Elissa, ``Title of paper if known,'' unpublished.
\bibitem{b5} R. Nicole, ``Title of paper with only first word capitalized,'' J. Name Stand. Abbrev., in press.
\bibitem{b6} Y. Yorozu, M. Hirano, K. Oka, and Y. Tagawa, ``Electron spectroscopy studies on magneto-optical media and plastic substrate interface,'' IEEE Transl. J. Magn. Japan, vol. 2, pp. 740--741, August 1987 [Digests 9th Annual Conf. Magnetics Japan, p. 301, 1982].
\bibitem{b7} M. Young, The Technical Writer's Handbook. Mill Valley, CA: University Science, 1989.
\end{thebibliography}

\end{document}